\documentclass[preprint2]{aastex}

\usepackage{graphicx}

\begin{document}

\title{New Solar Composition: The Problem With Solar Models Revisited}

\author{Aldo M. Serenelli}
\affil{Max  Planck Institute  for  Astrophysics, Karl  Schwarzschild Str.   1,
  Garching, D-85471, Germany} 

\author{Sarbani Basu}
\affil{Department of  Astronomy, Yale University, P.O. Box  208101, New Haven,
  CT 06520-8101, USA}

\author{Jason W. Ferguson}
\affil{Department of Physics,  Wichita State University, Wichita KS67260-0032,
  USA} 

\and

\author{Martin Asplund} 
\affil{Max  Planck Institute  for  Astrophysics, Karl  Schwarzschild Str.   1,
  Garching, D-85471, Germany} 

\begin{abstract}
We construct  updated solar  models with different  sets of  solar abundances,
including  the most  recent determinations  by  \citet{agss09}.
The latter work 
predicts  a  larger  ($\sim  10\%$)  solar metallicity  compared  to  previous
measurements 
by the same authors but significantly lower ($\sim 25\%$) than the recommended
value from a decade ago by \citet{gs98}. We compare the 
results  of our  models with  determinations of  the solar  structure inferred
through helioseismology measurements. The model that uses the most recent
solar  abundance determinations  predicts  the base  of  the solar  convective
envelope  to be located  at $R_{\rm  CZ}= 0.724{\rm  R_\odot}$ and  a surface
helium mass fraction of $Y_{\rm surf}=0.231$.  These results are in 
conflict with  helioseismology data ($R_{\rm  CZ}= 0.713\pm0.001{\rm R_\odot}$
and  $Y_{\rm  surf}=0.2485\pm0.0035$)  at  5$-\sigma$ and  11$-\sigma$  levels
respectively.   Using the new solar abundances,
we calculate the magnitude by which radiative opacities should be modified in
order to restore agreement with helioseismology. We find that a maximum change
of $\sim 15\%$  at the base of  the convective zone is required  with a smooth
decrease towards the core, where the change needed is $\sim 5\%$. The required
change  at  the base  of  the  convective envelope  is  about  half the  value
estimated previously. We  also present the solar neutrino  fluxes predicted by
the new  models. The  most important  changes brought about  by the  new solar
abundances  are the  increase by  $\sim 10\%$  in the  predicted  $^{13}$N and
$^{15}$O  fluxes  that arise  mostly  due  to the  increase  in  the  C and  N
abundances in the newly determined solar composition.

\end{abstract}


\maketitle

\section{Introduction}

Since the solar surface heavy-element content has been revised downwards by
~\citet[AGS05]{ags05}, from $(Z/X)_{\odot}=0.0229$ 
\citep[GS98]{gs98} to $(Z/X)_{\odot}=0.0165$, the excellent agreement 
between   standard   solar  model   (SSM)   predictions  and   helioseismology
determinations of the solar  structure \citep{model-S,bp00} has been seriously
compromised \citep{basu04,montalban,bs05}. 
This discrepancy between models and
helioseismic  inferences  has prompted  a  number  of  authors to  revise  the
physical inputs of SSMs 
\citep{montalban,basu04,bbps05,guzik06,jcd09}, and to question the revision of
the  solar  abundances,  particularly those  of  C,  N,  O,  Ne and  Ar  whose
fractional abundances cannot be determined from meteoritic samples
\citep{ab05,ab06,bbs05,dp06,ba08}. 

Very recently, \citet[hereafter AGSS09]{agss09}  have done a complete revision
of the solar photospheric abundances  for nearly all elements, including a new
3D hydrodynamical solar atmosphere  model with improved radiative transfer and
opacities 
\citep{tramp09}.  The predictions from this 3D model have been shown to agree 
remarkably  well   with  various  observational   constraints,  including  the
atmospheric 
thermal  structure  as  judged  from continuum  center-to-limb  variation  and
detailed line profile shapes 
\citep{pereira09a,pereira09b}. The newly determined solar abundances lead to 
$(Z/X)_{\odot}=0.0178$, higher than the AGS05 value but still well below older
determinations, e.g. GS98.

In  this Letter  we  present a  series  of new  SSM  calculations using  solar
compositions from  GS98, AGS05, and  the newly determined solar  abundances by
AGSS09.  All the  models incorporate new refinements in  the input physics, so
that models presented here with  the older compositions  (GS98 and AGS05)
represent   updated   versions    of   previous   solar   model   calculations
\citep{bs05}. For  each model, we compare our  results with helioseismological
determinations of solar properties and  also give the predicted solar neutrino
fluxes.  Additionally, and motivated by the new AGSS09 composition, we 
determine  the factor  by which  radiative opacities  in the
solar  interior should  be  increased  to solve  the  solar abundance  problem
following the scheme presented by \citet{jcd09}.

\section{Calculations and Results} \label{sec:calc}

Solar  models in  this work  have  been computed  with a  modified version  of
GARSTEC \citep{garstec}  that uses the nuclear energy  generation routine {\em
  exportenergy.f}\footnote{Publicly                available                at
  http://www.sns.ias.edu/$\sim$jnb}.  Element diffusion  in the solar interior
is  included according  to \citet{tbl94}.   Radiative opacities  are  from the
Opacity 
Project,    complemented    at    low    temperatures    with    those    from
\citet{lowt}. Specific  sets of opacities have  been computed for  each of the
solar compositions used in this paper (see below). 
With respect to previous works,
e.g. \citet{bs05,montecarlo}, the changes in  the input physics are: a revised
version           of          the          OPAL           equation          of
state\footnote{http://adg.llnl.gov/Research/OPAL/EOS\_2005/. 
OPAL  EOS  uses relative  metal  abundances  for C,  N,  O,  and  Ne close  to
\citet{gn93} and abundances of heavier elements are added to Ne. Differences in
relative abundances  of metals with  more recent solar  abundance compilations
have negligible influence  in the global properties of  solar models, provided
the correct overall metallicity is used \citep{gong01,opal01}.}  (EOS) that
corrects errors in the 2002  OPAL EOS tables from \citet{opal01} (our previous
choice), and  updated values of  two important nuclear  astrophysical factors,
S$_{34}$ \citep{s34} and S$_{1,14}$ \citep{s114}, the latest determinations by
the LUNA experiment.

We have  computed solar models  using three different basic  solar abundances.
Two models employ previous solar abundance 
compilations  (GS98 and  AGS05) and  show  small differences  with respect  to
models with the same  abundances presented elsewhere, e.g.  \citet{bs05}. The
changes  originate from  the use  of the  updated EOS  and cross-sections for
nuclear reactions mentioned above. A  third SSM has been computed adopting the
new solar composition determined by AGSS09. The most important results in this
work are related to this model.  The choice  of the  abundance scale
(meteoritic or photospheric) deserves a short discussion.  
While AGSS09  find the average difference between  photospheric and meteoritic
abundances  to be $0.00  \pm 0.04$~dex,  a few  elements relevant  to detailed
solar modeling  show comparable  or slightly larger  deviations. This  is the
case for Mg, Ca, and Fe for which differences between the two scales are 0.07,
0.05, and  0.05~dex respectively, photospheric values being  larger. Given the
historical  robustness and  higher  accuracy of  meteoritic determinations  of
abundance   ratios,  and   the  present   excellent  overall   agreement  with
photospheric  abundances,  we  adopt  for  AGSS09  meteoritic  abundances  for
refractory elements as the standard choice for our solar models. We also
investigate,   however,  the   use   of  adopting   the  photospheric   values
throughout. This is also 
consistent with the  adoption of meteoritic scales in  previous works on solar
models  (e.g.  \citealp{bs05,montecarlo}). The  abundances  for the  different
solar compositions used  in this paper are given  in Table~\ref{tab:compo}, in
particular for those elements entering the calculation of radiative opacities.

Models have been  evolved from the pre-main sequence  to the present-day solar
system age,  $\tau_\odot=4.57~\mbox{Gyr}$\footnote{Detailed structure of solar
  models   at  $\tau_\odot$   presented  in   this  work   can  be   found  in
  http://www.mpa-garching.mpg.de/$\sim$aldos}. The main characteristics of the
models are listed in Table~\ref{tab:ssm}. The second and 
third columns give  the present-day heavy elements to  hydrogen mass ratio and
the surface metallicity.   From the fourth to the  seventh columns we present
quantities 
directly related to helioseismology: surface helium mass fraction $Y_{\rm
surf}$, depth of the convective zone 
$R_{\rm CZ}/{\rm R_\odot}$, and the average rms difference between
model and solar sound speed and density profiles, $\left< \delta c/c \right>$
and $\left< \delta \rho/\rho \right>$ respectively. 
Columns eighth and ninth give the central helium mass fraction $Y_{\rm c}$ and 
metallicity $Z_{\rm c}$ at $\tau_\odot$.  Finally, the last three columns give
the initial composition of the models and the mixing length parameter.

Results for the  GS98 and AGS05 models are very similar  to those presented in
\citet{bs05}; the improved EOS leads to changes in the sound speed and density
profiles about  one order  of magnitude smaller  than the  differences between
solar  models and  helioseismic  inferences, while  changes  in nuclear  cross
sections only  affect predictions  for neutrino fluxes  that are  discussed in
detail below.  Helioseismically derived  values for $R_{\rm CZ}/{\rm R_\odot}$
and $Y_{\rm  surf}$ and are $0.713  \pm 0.001$ \citep{basu97}  and $0.2485 \pm
0.0035$  \citep{basu04}   respectively.   The  GS98   model  predicts  $R_{\rm
  CZ}/R_\odot$  with the right  value compared  to helioseismology,  while the
AGS05 model shows a 15$-\sigma$  discrepancy. For $Y_{\rm surf}$ the situation
is  analogous: GS98 value  differs from  the helioseismology  determination by
1.8-$\sigma$ while for AGS05 the discrepancy is 5.5-$\sigma$.

\begin{figure}
\includegraphics[scale=.55]{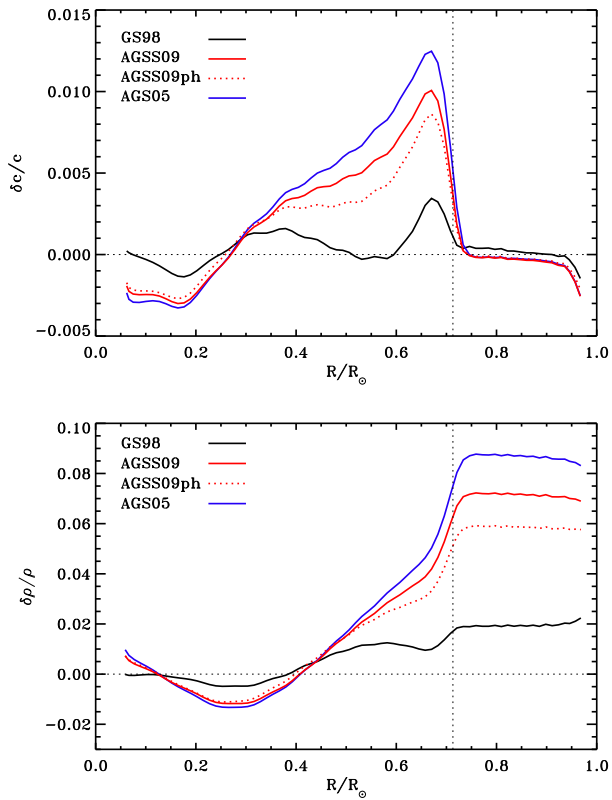}
\caption{Relative sound  speed $\delta c/c$  and density $\delta \rho  / \rho$
  differences,  in the  sense (Sun  - Model)/Model,  between solar  models and
  helioseismological  results. Details  on  the inversion  procedure and  data
  used, as well as the reference sound speeds and densities are given in
  \citet{bisoniii}. \label{fig:cs}} 
\end{figure}

For the SSM that adopts the newly determined solar composition AGSS09, 
$R_{\rm CZ}/{\rm  R_\odot}$ and $Y_{\rm surf}$
show some  improvements with  respect to the  AGS05 model, but
still far away from the helioseismology values by 
11-$\sigma$ and 5-$\sigma$ respectively.  The slight increase in oxygen 
abundance (0.03\,dex) and the larger  change in neon (0.09\,dex) contribute to
enhance 
the opacity  below the convective  zone (CZ), deepening its inner boundary
and decreasing the mismatch with the  solar sound speed at the same time. This
is illustrated  in the top panel of  Figure~\ref{fig:cs} where  the relative
difference in sound speed is shown for the models considered in this work. The
peak in  the profile  of the sound  speed difference,  right below the  CZ, is
$\sim 1\%$  for the AGSS09 composition,  an improvement with  respect to AGS05
but  still significantly  higher than  that for  the GS98  model.  The overall
agreement with 
the  solar sound  speed, as  derived by  inversions and  indicated  by $\left<
\delta c/c\right>$, is a factor of four times worse for the AGSS09 model than 
for GS98. For the latter model, $\left< \delta c/c\right>$ is almost unchanged 
compared to results from previous works. However, 
a detailed comparison of the sound speed profiles shown in Figure~\ref{fig:cs}
with those  presented in \citet{bs05}  unveils some differences,  more evident
below  $R \approx 0.6$\,R$_\odot$.   These
changes are not related to the improved physics adopted in the new models, but
rather the result of using better  data for low-degree
($\ell \leq  3$) modes that  penetrate the solar core  (see \citealt{bisoniii}
for details). In the same figure, results for 
density inversions are shown in the bottom panel. Again, the AGSS09 composition
leads  to an  improvement in  the agreement  with helioseismology  compared to
the  AGS05  model, but  still  far  from the  results  obtained  for the  GS98
composition.  

As already  mentioned, meteoritic and photospheric abundances  in AGSS09 agree
with each other very well, but  a few elements show differences that could have
potential  impact  on  details  in  the solar  structure.   To  quantify  this
assertion,  we have  computed an  additional SSM  using only  the photospheric
abundances  given  in  AGSS09,  for  which  $(Z/X)_{\odot}=0.0181$.  The  main
characteristics of this model, identified as AGSS09ph, are given in the
last entry of  Table~\ref{tab:ssm}. Compared to the model  with the meteoritic
abundances, AGSS09ph performs somewhat better in terms of helioseismological
quantities  as  inferred  from  the  results summarized  in  the  table,  with
discrepancies with the measured depth of the CZ and surface helium abundance 
of the order of 9-$\sigma$ and 4-$\sigma$ respectively. 
The  sound speed  and density  profiles are shown  as dotted  lines in
Figure~\ref{fig:cs}.   The  changes  with   respect  to  our  standard  AGSS09
(meteoritic scale) model changes are
mostly due to the larger Mg  and Fe photospheric abundances (0.07 and 0.05~dex,
respectively) that enhance the
opacity  in  the  radiative  interior;  the  fractional  increase  in
opacity  is larger  than the  fractional increase  in the  overall metallicity
(note the largest improvement in the sound speed, for instance, occurs at
$R \approx 0.5$\,R$_\odot$, the region where the contribution of Mg to the
opacity is largest). Our results  show that adoption of the photospheric scale
gives  slight   improvements  in   the  solar  model   predictions.   
However, since the uncertainties in the meteoritic abundances typically are
smaller than the corresponding ones for the photospheric determinations
the meteoritic scale is our preferred choice for solar
abundances (with the exception of the volatile elements that are depleted
in meteorites).  This  is reinforced by the  historical
robustness of meteoritic abundance determinations.

\begin{figure}
\includegraphics[scale=.55]{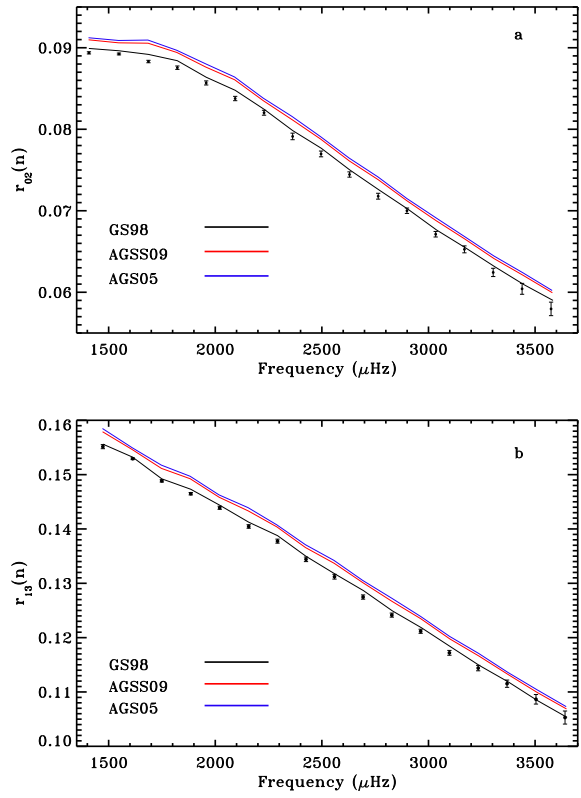}
\caption{Separation  ratios. Comparison between  values determined  from BiSON
  data and the solar models presented in this work.\label{fig:seprat}}
\end{figure}

Low-degree  helioseismology ($\ell  \leq 3$)  can  be used  to derive  seismic
information on the solar core,  particularly by using the so-called separation
ratios as described by  \citet{roxb}. \citet{bisonii} have used the separation
ratios constructed with very  precise frequencies of low-$\ell$ modes measured
by the  Birmingham Solar-Oscillations Network (BiSON)  to constrain properties
of the  solar core.  They showed the  discrepancy in solar  models constructed
with AGS05  composition extends all the  way to the  solar core and it  is not
just related to deficiencies in the modeling of the solar outer layers, in the
convective envelope. 
Here we compare the observed separation ratios to those computed for our solar 
models;  results  are  displayed  in Figure~\ref{fig:seprat}.  As  with  other
helioseismology indicators, the GS98 model performs much better than the AGS05
model.  The adoption  of the  AGSS09  composition in  the SSM  has very  small
influence in the  core structure of the model, as  it practically overlaps the
AGS05 model.  Results  for the AGSS09ph  model closely resemble  those from
AGSS09 and, for clarity, are not shown in Figure~\ref{fig:seprat}.
As discussed in \citet{bisoni}, 
values  of  the  separation  ratios   are  closely  related  to  the  quantity
$1/r  \ (dc/dr)$  integrated over  the  solar structure.  Differences in  this
quantity between
models with  AGS05 and AGSS09 (both meteoritic  and photospheric) compositions
are  very   small  and  only   present  very  close   to  the  center   ($r  <
0.05~R_\odot$) and thus have a very small impact on the separation ratios.

Solar neutrino fluxes  for the models are listed  in Table~\ref{tab:neu}.  The
new  astrophysical factor  $S_{34}$ \citep{s34}  is 7\%  larger  than previous
determination and it is 
responsible for  the larger $^7$Be  and $^8$B fluxes  of models GS98  and AGS05
with  respect to  those published  in \citet{montecarlo}  with the  same solar
compositions.  Similarly, the somewhat smaller (7.6\%) $S_{1,14}$ value recently
published  by the  LUNA  collaboration \citep{s114}  leads  to a  proportional
reduction in the $^{13}$N and  $^{15}$O fluxes. The increase in metallicity in
the AGSS09 solar composition, 
compared to AGS05, is only partially  reflected in the changes of the neutrino
fluxes. This is  so because elements that more strongly  affect the solar core
temperature  (Si,  S,  and Fe)  have  the  same  abundance  in AGS05  and  the
meteoritic AGSS09 scale.  The increase by 0.09~dex in Ne and by 0.22~dex in
Ar are the most important changes in abundances influencing the core
temperature. This is reflected, for instance, in the $\sim 4\%$ increase in the
$^8$B flux  from AGS05  to AGSS09. $^{13}$N  and $^{15}$O fluxes,  that depend
almost  linearly in the  added abundance  of carbon  and nitrogen  show larger
changes (of the order of $\sim 12 - 14\%$) due to the 0.03 and 0.05~dex larger
abundances of these elements in  the new AGSS09 abundances. Differences in the
neutrino fluxes  between the AGSS09  and the GS98  models are of the  order of
10\% for $^7$Be,  20\% for $^8$B and 38\% for the  added $^{13}$N and $^{15}$O
fluxes. For  the sake of completeness  we present the neutrino  fluxes for the
AGSS09ph  model.  The  effect  of  the increased  iron  abundance  is  readily
noticeable particularly in the larger $^8$B flux compared to AGSS09 model and,
to a lesser extent in the CNO and $^7$Be fluxes. 
The  possibilities  that current  and  future  neutrino  experiments offer  to
constrain the  solar core  metallicity are  beyond the
scope of this paper and are discussed elsewhere \citep{cn,bps}.

Qualitatively, the  AGSS09 abundances do  not change the picture  that emerged
with  the previous  set  of  solar abundances,  AGS05:  SSMs constructed  with
abundances  derived from the  most sophisticated  solar atmosphere  models and
up-to-date atomic  data conflict with all helioseismology  inferences of solar
structure.   Potential solutions to  the solar  abundance problem  analyzed by
different authors, none of them  successful in restoring the agreement between
SSM  and helioseismology  measurements, still  face the  same problems  if the
AGSS09  composition  is  used  instead   of  AGS05.  In  this  regard,  it  is
particularly informative the  work by \citet{dp06}, which show  in the $R_{\rm
CZ}-Y_{\rm surf}$ plane  the direction in which model  predictions change when
modifications in their input physics are applied. It is not a simple
task to  find deficits  in the models  that produce simultaneously  changes in
both  $R_{\rm CZ}$  and  $Y_{\rm surf}$  in  the right  direction, except  for
restoring  the metallicity  to a  larger value,  comparable to  that  of GS98.
Quantitatively,  however,   the  disagreement  is  less   severe  with  AGSS09
composition and, motivated by this fact, it
is worth  taking another  look at  the opacity deficit  induced by  the AGSS09
abundances.  We  have  done  a  similar  analysis  to  the  one  presented  by
\citet{jcd09} to  check by how much opacities  in the AGSS09 model  have to be
increased to recover the level of agreement with helioseismology that the GS98
model gives.  We find that in the central regions the required change is $\sim
5\%$  (2\%  with AGSS09ph),  very  close  to  what \citet{jcd09}  found.   The
magnitude of  the change increases  smoothly outwards and reaches  $\sim 15\%$
(12\% with AGSS09ph)  at the bottom of the CZ. This  requirement is smaller by
almost a factor of two than that found by 
\citet{jcd09},  very likely  because they  used  the S-model  (which used  the
higher $(Z/X)_\odot=0.0245$ from \citealp{gn93})  as their reference model and
one with the AGS05 solar composition,  while we have used GS98 and AGSS09 (see
\citealp{sere10} for a more thorough discussion). We
note the  required changes are much  larger than differences  found between OP
and OPAL opacities  in the radiative solar interior.  It would be interesting,
as  pointed  out by  \citet{jcd09}, to find  other
observable implications that changes of  $12-15$\% in the radiative opacity at
temperatures of a couple to a few millon degrees would have, independently
of those from solar models.  

\section{Summary} \label{sec:summ}

We  have  computed  new  standard  solar  models  that  incorporate  the  most
up-to-date input physics, including the updated OPAL equation of state and the
most  recent determinations  of the  astrophysical factors  for  the important
$^3{\rm He}(^4{\rm He},\gamma)^7$Be and 
$^{14}{\rm N}({\rm  p},\gamma)^{15}$O reactions. We have  used three different
sets of solar abundances: GS98, AGS05 and the newly determined AGSS09. We have
found that the updated physical inputs have very little effect on the 
properties  of solar  models, both  in  terms of  solar structure  and in  the
neutrino fluxes.  
The most important results in this work are those from the model that adopts
the new set of solar abundances, AGSS09. The new abundances are determined
by  using an  improved solar  atmosphere  model and  atomic data,  and a  more
careful selection  of spectroscopic lines. The resulting  solar metallicity is
somewhat   larger  than   that  from   AGS05.   This  is   reflected  in   the
helioseismological properties  of the solar  model. For the AGSS09  model, the
sound speed and  density profiles, the predicted surface  helium abundance and
the   depth  of   the  convective   zone  are   still  in   conflict  with
helioseismology data,  although the disagreement  is less severe than  for the
AGS05 model.  Still, results are far  from the excellent match  found with the
GS98 composition. Finally, we have found that if radiative opacities 
were to be modified to restore the agreement between solar models (with AGSS09
composition) and  helioseismology, the  required changes are  
$\approx 12-15$\% right  below the convective zone  with a smooth  decrease towards the
central regions, where changes should be $2-5$\%.

\acknowledgements AMS thanks Maria Bergemann for insightful discussions on 
the physics of non-LTE effects in the formation of spectral lines. 
SB is  partially supported  by NSF grant  ATM-0348837. We thank  the anonymous
referee for useful suggestions and comments.

\clearpage

\begin{deluxetable}{lcccc}
\tablewidth{0pt}
\tablecaption{Adopted solar chemical compositions. \label{tab:compo}} 
\tablehead{\colhead{} & \multicolumn{4}{c}{$\log{\epsilon}$} \\ 
\cline{2-5}
\\
\colhead{Element} & \colhead{GS98}  & \colhead{AGS05$^a$} & \colhead{AGSS09$^a$}
& \colhead{AGSS09ph$^b$}}\startdata
C & 8.52 & 8.39 & 8.43 & 8.43 \\
N & 7.92 & 7.78 & 7.83 & 7.83 \\
O & 8.83 & 8.66 & 8.69 & 8.69 \\
Ne & 8.08 & 7.84 & 7.93 & 7.93 \\
Na & 6.32 & 6.27 & 6.27 & 6.24 \\
Mg & 7.58 & 7.53 & 7.53 & 7.60 \\
Al & 6.49 & 6.43 & 6.43 & 6.45 \\   
Si & 7.56 & 7.51 & 7.51 & 7.51 \\
S & 7.20 & 7.16 & 7.15 & 7.12 \\ 
Ar & 6.40 & 6.18 & 6.40 & 6.40 \\ 
Ca & 6.35 & 6.29 & 6.29 & 6.34 \\ 
Cr & 5.69 & 5.63 & 5.64 & 5.64 \\
Mn & 5.53 & 5.47 & 5.48 & 5.43 \\
Fe & 7.50 & 7.45 & 7.45 & 7.50 \\
Ni & 6.25 & 6.19 & 6.20 & 6.22 \\
\enddata
\tablecomments{Abudances given  as $\log{\epsilon_i}\equiv\log{N_i/N_H}+12$.\\
  $^a$ The adopted abundances are the recommended solar photospheric abundances
  for  the volatile  elements  (C, N,  O, Ne  and  Ar) and  the CI  chondritic
  meteoritic values for the 
  remaining elements. 
  $^b$  The   adopted  abundances  are  the   recommended  solar  photospheric
  abundances throughout. } 
\end{deluxetable}

\clearpage

\begin{deluxetable}{lccccccccccc}
\tablewidth{0pt}
\tablecaption{Main characteristics of solar models.\label{tab:ssm}}
\tablehead{Model  & $(Z/X)_{\rm  surf}$ &  $Z_{\rm surf}$  & $Y_{\rm  surf}$ &
  $R_{\rm CZ}/R_\odot$ & 
$\left< \delta  c / c\right>$ &  $\left< \delta \rho /  \rho\right>$ & $Y_{\rm
    c}$ & $Z_{\rm c}$ & 
$Y_{\rm ini}$ & $Z_{\rm ini}$
& $\alpha_{\rm MLT}$}
\startdata
GS98 & 0.0229 &  0.0170 & 0.2423 & 0.713 & 0.0010 & 0.011  & 0.6330 & 0.0201 &
0.2721 & 0.0187 & 2.15 \\ 
AGS05 & 0.0165 & 0.0126 & 0.2292 &  0.728 & 0.0049 & 0.048 & 0.6195 & 0.0149 &
0.2593 & 0.0139 & 2.10 \\ 
AGSS09 & 0.0178 & 0.0134 & 0.2314 & 0.724 & 0.0038 & 0.040 & 0.6220 & 0.0160 &
0.2617 & 0.0149 & 2.09 \\ 
AGSS09ph & 0.0181 & 0.0136 & 0.2349 & 0.722 & 0.0031 & 0.033 & 0.6263 & 0.0161
& 0.2653 & 0.0151 & 2.12 \\ 
\enddata
\end{deluxetable}

\clearpage

\begin{deluxetable}{lcccccccc}
\tablewidth{0pt}
\tablecaption{Predicted neutrino fluxes. \label{tab:neu}}
\tablehead{Model &  pp & pep &  hep & $^7$Be &  $^8$B & $^{13}$N  & $^{15}$O &
  $^{17}$F} 
\startdata
GS98 & 5.97 & 1.41 & 7.91 & 5.08 & 5.88 & 2.82 & 2.09 & 5.65 \\
AGS05 & 6.04 & 1.44 & 8.24 & 4.54 & 4.66 & 1.85 & 1.29 & 3.14 \\
AGSS09 & 6.03 & 1.44 & 8.18 & 4.64 & 4.85 & 2.07 & 1.47 & 3.48 \\
AGSS09ph & 6.01 & 1.43 & 8.10 & 4.79 & 5.22 & 2.15 & 1.55 & 3.70 \\
\enddata
\tablecomments{Neutrino   fluxes  are   given  in   units   of  $10^{10}$(pp),
  $10^9$($^7$Be), $10^8$(pep, $^{13}$N, $^{15}$O), $10^6$($^8$B, $^{17}$F) 
and $10^3$(hep)~\hbox{cm$^{-2}$ s$^{-1}$}.  Direct measurement 
by the Borexino experiment of the $^7$Be
flux  gives $5.18 \pm  0.51 \times  10^9$~\hbox{cm$^{-2}$ s$^{-1}$}  after 192
days  of  data \citep{borex}.  For  $^8$B,  and  until the  SNO  collaboration
presents a combined analysis of the three phases of the experiment,
a simple weighed average of the three phases of
the  SNO   experiment  \citep{sno1,sno2,sno3}  gives  $5.18   \pm  0.29  \times
10^6$~\hbox{cm$^{-2}$ s$^{-1}$} for this flux.} 
\end{deluxetable}

\clearpage

\clearpage

\begin{thebibliography}{}

\bibitem[Ahmad et al.(2002)]{sno1} Ahmad, Q.~R., et al.\ 
2002, Physical Review Letters, 89, 011301 

\bibitem[Aharmim et al.(2005)]{sno2} Aharmim, B., et al.\ 
2005, \prc, 72, 055502

\bibitem[Aharmim et al.(2008)]{sno3} Aharmim, B., et al.\ 
2008, Physical Review Letters, 101, 111301

\bibitem[Antia \&  Basu(2005)]{ab05} Antia, H.~M.,  \& Basu, S.\  2005, \apjl,
620, L129

\bibitem[Antia  \& Basu(2006)]{ab06} Antia,  H.~M., \&  Basu, S.\  2006, \apj,
644, 1292

\bibitem[Arpesella et al.(2008)]{borex} Arpesella, C., et 
al.\ 2008, Physical Review Letters, 101, 091302

\bibitem[Asplund et al.(2005)]{ags05} Asplund, M., Grevesse, N., \& Sauval, J.
  2005, Cosmic Abundances as Records of Stellar Evolution and Nucleosynthesis,
  336, 25

\bibitem[Asplund et al.(2009)]{agss09} Asplund,  M., Grevesse, N., Sauval, J.,
  \& Scott, P. 2009, \araa, 47, 481

\bibitem[Badnell  et al.(2005)]{op} Badnell,  N.~R., Bautista,  M.~A., Butler,
K., Delahaye, F., Mendoza, C.,  Palmeri, P., Zeippen, C.~J., \& Seaton, M.~J.\
2005, \mnras, 360, 458

\bibitem[Bahcall et al.(2005)]{bbps05} Bahcall, J.~N., Basu, S., Pinsonneault,
M., \& Serenelli, A.~M.\ 2005, \apj, 618, 1049

\bibitem[Bahcall et al.(2005)]{bbs05} Bahcall,  J.~N., Basu, S., \& Serenelli,
A.~M.\ 2005, \apj, 631, 1281

\bibitem[Bahcall  et  al.(2001)]{bp00} Bahcall,  J.~N.,  Pinsonneault, M.,  \&
Basu, S. 2001, \apj, 555, 990

\bibitem[Bahcall et  al.(2006)]{montecarlo} Bahcall, J.~N.,  Serenelli, A.~M.,
\& Basu, S.\ 2006, \apjs, 165, 400

\bibitem[Bahcall  et  al.(2005)]{bs05} Bahcall,  J.~N.,  Serenelli, A.~M.,  \&
Basu, S.\ 2005, \apjl, 621, L85

\bibitem[Bahcall  et  al.(2004)]{cz}  Bahcall,  J.~N.,  Serenelli,  A.~M.,  \&
Pinsonneault, M.\ 2004, \apj, 614, 464

\bibitem[Basu \& Antia(1997)]{basu97} Basu, S., \& Antia, H.~M.\ 1997, \mnras,
287, 189

\bibitem[Basu \& Antia(2004)]{basu04}  Basu, S., \& Antia, H.  M. 2004, \apjl,
606, L85

\bibitem[Basu \& Antia(2008)]{ba08} Basu, S., \& Antia, H.~M.\ 2008, \physrep,
457, 217

\bibitem[Basu et  al.(2007)]{bisoni} Basu,  S., Chaplin, W.~J.,  Elsworth, Y.,
  New, R., Serenelli, A.~M., \& Verner, G.~A.\ 2007, \apj, 655, 660

\bibitem[Basu et al.(2009)]{bisoniii} Basu,  S., Chaplin, W.~J., Elsworth, Y.,
New, R., \& Serenelli, A.~M.\ 2009, \apj, 699, 1403

\bibitem[Chaplin  et  al.(2007)]{bisonii}  Chaplin, W.~J.,  Serenelli,  A.~M.,
Basu, S., Elsworth, Y., New, R., \& Verner, G.~A.\ 2007, \apj, 670, 872

\bibitem[Christensen-Dalsgaard  et  al.(1996)]{model-S} Christensen-Dalsgaard,
  J. C., et al.\ 1996, Science, 272, 1286

\bibitem[Christensen-Dalsgaard et al.(2009)]{jcd09} Christensen-Dalsgaard, J.,
di Mauro, M.~P., Houdek, G., \& Pijpers, F.\ 2009, \aap, 494, 205

\bibitem[Costantini et  al.(2008)]{s34} Costantini, H., et  al.\ 2008, Nuclear
Physics A, 814, 144

\bibitem[Delahaye \& Pinsonneault(2006)]{dp06}  Delahaye, F., \& Pinsonneault,
M.~H.\ 2006, \apj, 649, 529

\bibitem[Ferguson  et  al.(2005)]{lowt}  Ferguson,  J.~W.,  Alexander,  D.~R.,
Allard, F., Barman, T.,  Bodnarik, J.~G., Hauschildt, P.~H., Heffner-Wong, A.,
\& Tamanai, A.\ 2005, \apj, 623, 585

\bibitem[Gong et al.(2001)]{gong01} Gong, Z., D{\"a}ppen, W., 
\& Nayfonov, A.\ 2001, \apj, 563, 419

\bibitem[Grevesse  \& Noels(1993)]{gn93}  Grevesse,  N., \&  Noels, A.\  1993,
Origin and Evolution of the Elements, 15

\bibitem[Grevesse \& Sauval(1998)]{gs98} Grevesse, N., \& Sauval, A.~J.\ 1998,
Space Science Reviews, 85, 161

\bibitem[Guzik(2006)]{guzik06} Guzik, J.~A.\ 2006, Proceedings of SOHO 18/GONG
2006/HELAS I, Beyond the spherical Sun, 624, 17

\bibitem[Haxton  \& Serenelli(2008)]{cn} Haxton,  W.~C., \&  Serenelli, A.~M.\
2008, \apj, 687, 678


\bibitem[Marta et al.(2008)]{s114} Marta, M., et al.\ 2008, \prc, 78, 022802


\bibitem[Montalb{\'a}n et~al.(2004)]{montalban} Montalb{\'a}n, J., Miglio, A.,
Noels, A., Grevesse, N., \& di  Mauro, M.~P. 2004, in ESA Special Publication,
Vol.  559, SOHO  14  Helio-  and Asteroseismology:  Towards  a Golden  Future,
ed. D.~Danesy, 574--+

\bibitem[Pe\~na-Garay \& Serenelli(2008)]{bps} Pe\~na-Garay, C., \& Serenelli,
  A.\ 2008, arXiv:0811.2424

\bibitem[Pereira et al.(2009a)]{pereira09a}  
Pereira, T.~M.~D.,  Asplund, M.,  Trampedach, R., \&  Collet, R.\  2009, \aap,
submitted 

\bibitem[Pereira et al.(2009b)]{pereira09b}  
Pereira, T.~M.~D., Kiselman, D., \& Asplund, M.\ 2009, \aap, submitted 

\bibitem[Rogers  \&   Nayfonov(2002)]{opal01}  Rogers,  F.~J.,   \&  Nayfonov,
  A.\ 2002, \apj, 576, 1064

\bibitem[Roxburgh  \& Vorontsov(2003)]{roxb}  Roxburgh,  I.~W., \&  Vorontsov,
S.~V.\ 2003, \aap, 411, 215

\bibitem[Serenelli(2010)]{sere10} Serenelli, A.  2010, \apss \ in Synergies
Between Solar and  Stellar Modelling, ed. M.  Marconi, D.  Cardini \& M. P.
  Di Mauro, in preparation
 
\bibitem[Thoul et al.(1994)]{tbl94} Thoul, A.~A., Bahcall, 
J.~N., \& Loeb, A.\ 1994, \apj, 421, 828

\bibitem[Trampedach et al.(2009)]{tramp09} Trampedach, R., Asplund, M., Hayek,
W., \& Collet, R. 2009, \aap, to be submitted

\bibitem[Weiss \&  Schlattl(2008)]{garstec} Weiss, A., \&  Schlattl, H.\ 2008,
\apss, 316, 99

\end{thebibliography}
\end{document}